\documentclass[12pt]{article}
\usepackage{amsfonts,amsbsy}
\newcommand{\be}{\begin{equation}}
\newcommand{\ee}{\end{equation}}
\newcommand{\ba}{\begin{array}}
\newcommand{\ea}{\end{array}}
\newcommand{\p}{\partial}

\DeclareMathAlphabet{\bi}{OML}{cmm}{b}{it}
\def\dder#1{\p_{#1}}

\def\eqref#1{{\rm(\ref{#1})}}

\date{July 11, 2011}

\author{{\sc M. Marvan and A. Sergyeyev}\\
Mathematical Institute,
Silesian University in Opava,\\ Na Rybn\'\i{}\v cku 1, 746\,01 Opava,\\
Czech Republic\\
E-mail: {\tt Michal.Marvan@math.slu.cz},
\tt{Artur.Sergyeyev@math.slu.cz}}
\begin{document}
\title{\protect\vspace*{-15mm}\Large\bf Recursion operators
for dispersionless integrable systems in any dimension\protect\vspace{-3mm}}
\maketitle

\begin{abstract}\protect\vspace*{-15mm}
\centering
    \begin{minipage}{0.9\textwidth}
We present a new approach to construction of recursion operators
for multidimensional integrable systems which
have a Lax-type representation in
terms of a pair of commuting vector fields.
It is illustrated by the examples
of the Manakov--Santini system which is a hyperbolic
system in $N$ dependent and $N+4$ independent variables, where
$N$ is an arbitrary natural number,
the six-dimensional generalization of the
first heavenly equation, the modified heavenly equation,
and the dispersionless Hirota equation.
\looseness=-1
\end{minipage}
\end{abstract}


\section{Introduction}

Existence of an infinite hierarchy of local or nonlocal symmetries usually indicates
integrability of a PDE system.
A standard tool to produce such hierarchies is a recursion operator, see e.g.\
\cite{Bl,olv_eng2,Z-K}. In essence, if one has an auxiliary linear system and a class of its coefficients for which the inverse scattering problem can be solved, 
then using a recursion operator
it is usually possible to write down the most general nonlinear system whose solutions
can be obtained using the inverse scattering problem in question, cf.\ e.g.\ 
\cite{akns1, N}.

Recursion operators can be sought for in a variety of ways, depending on the definition
used. For instance, a typical recursion operator in (1+1) dimensions is a pseudodifferential operator which maps symmetries into symmetries \cite{O} and
can be derived from the Lax pair, see e.g.\ \cite{O, G-K-S, K} and references therein.
However, these definitions and methods do not immediately extend to higher dimensions~\cite{Z-K}. Instead, a higher-dimensional equation may admit a bilocal recursion
operator as introduced by Fokas and Santini, see e.g.~\cite{F-S,S-F1,F-S2}, and  \cite{K2};
a prototypical example here is the Kadomtsev--Petviashvili equation.\looseness=-1

From a different perspective,
recursion operators are B\"acklund auto-transformations of a linearized equation.
They first appeared in this form in several works by Papachristou~\cite{Pap1,Pap2,Pap3}.
The idea is that symmetries are essentially solutions of the linearized equation
and B\"acklund transformations are the most general transformations that relate solutions to solutions.

However, as pointed out by one of us~\cite{Mar}, the same point of view applies to
Guthrie's \cite{Gut1} generalized recursion operators in (1+1) dimensions.
Guthrie introduced them to avoid difficulties connected with
the lack of rigorous interpretation of action of pseudodifferential
operators on symmetries.

The recursion operators which are auto-B\"acklund transformations
of linearized systems appear to exist only for
a certain class of integrable systems, which has yet to be
characterized in full generality. Such recursion
operators are closely related to
zero curvature representations whenever the latter exist,
see e.g.~\cite{MaSe} and references
therein.

By a zero curvature representation for a system of PDEs $F=0$ we mean
a one-form $\alpha =\sum_i A_i\,dx^i$ which satisfies
$D_{x^j} A_i - D_{x^i} A_j + [A_i, A_j] = 0$
on the solution manifold of $F=0$;
here $x^i$ are the independent variables and
$A_i$ belong to some matrix Lie algebra.
Then the operators $D_{x^i} - \mathrm{ad} A_i$
commute and we can define a matrix pseudopotential $\Psi$ by setting
\begin{equation}\label{Psi}
\Psi_{x^i} - [A_i,\Psi] = \partial_{\bi{U}} A_i,
\end{equation}
where $\partial_{\bi{U}}$ denotes linearization
along a symmetry $\bi{U}$ (see Section 2 below for details).

For the majority of integrable systems in (1+1) dimensions
the pseudopotentials $\Psi$ provide
nonlocal terms of inverse recursion operators, whereas
local terms thereof are, as a rule, limited to zero-order ones at most.
This approach applies to multidimensional systems
whenever a zero curvature representation is available.
For instance, the recursion operator found by Papachristou
in \cite{Pap1} is easily seen to be of this kind.\looseness=-1
\looseness=-1

Now turn to dispersionless multidimensional systems which can be written as
a commutativity condition for a pair of first-order
linear scalar differential operators with no free terms
(i.e., vector fields) and no derivatives with respect to the spectral parameter;
see e.g.\ \cite{Z-S, M-S, B-K, M-S2, D} and references therein
for more information on such systems. The systems of this kind are
a subject of intense research as they arise in a multitude of areas from
self-dual gravity, see e.g.\ \cite{G, ney, shef1, D},
hyper-K\"ahler \cite{GS}, symplectic \cite{df} and conformal \cite{bft} geometry
to fluid dynamics and related fields, cf.\ e.g.\ \cite{RS, F, FMS}.
Even though the recursion operators for some of these systems
were found, see e.g.\ \cite{Pap1, S, ney, M-S2, She, shef1},
they were obtained using either various {\em ad hoc} methods or the partner symmetry method, both of which can be applied only under fairly restrictive assumptions.
\looseness=-1

Below we present a method for finding recursion operators 
which is based on a generalization of pseudopotentials (\ref{Psi})
using the adjoint representation of the Lie algebra of vector fields.
We are convinced that our approach applies to a considerably broader 
class of dispersionless systems than the methods mentioned 
in the previous paragraph. Moreover, our method is also more algorithmic:
given a Lax-type representation for the system under study, 
finding a recursion operator of the type described in our paper, if it exists, 
is an essentially algorithmic task while 
e.g.\ the partner symmetry approach involves a non-algorithmic subproblem 
of representing the linearized equation as a two-dimensional divergence.

The paper is organized as follows.
In Section~2 we present a general
construction of recursion operators
which are auto-B\"acklund transformations
of linearized systems, and Sections~3 and~4 illustrate its application
on the examples of the Manakov--Santini system
and the dispersionless Hirota equation.
Finally, in Section~5 we give a modification of the construction
from Section~2 for the case of Hamiltonian
vector fields and provide some further examples.
\looseness=-1

\section{The general approach}\label{ga}
Let $F=0$ be a system of PDEs in $n$ independent variables $x^i$, $i=1,\dots,n$, for
the un\-known $N$-com\-ponent vector function
$\bi{u}=(u^1,\dots,u^N)^{T}$,
where the superscript `$T$' denotes matrix transposition.
Denote
$u^\alpha_{i_1\dots i_n}=\p^{i_1+\cdots+i_n}u^\alpha/\p (x^1)^{i_1}\cdots \p  (x^n)^{i_n}$;
in particular, $u^\alpha_{00\cdots0}\equiv u^\alpha$.
As usually in the formal theory of PDEs~\cite{olv_eng2,vinbook,Bl},
$x^i$ and $u^\alpha_{i_1\dots i_n}$ are considered as independent quantities and can be
viewed as coordinates on an abstract infinite-dimensional space (a jet space).
By a {\it local function} or a function on the jet space we shall mean any function of a
finite number of $x^i$ and $\bi u$ and their derivatives.
We denote
$$
D_{x^j} = \frac{\p}{\p x^i}
 + \sum_{\alpha=1}^{N} \sum_{i_1,\dots,i_n=0}^{\infty}
   u^\alpha_{i_1\dots i_{j-1} (i_j+1) i_{j+1}\dots i_n} \frac{\p}{\p u^\alpha_{i_1\dots i_n}}
$$
the usual total derivatives, which can be naturally viewed as vector fields on the jet space.
The condition $F=0$ along with its differential consequences
$D_{x^1}^{i_1}\cdots D_{x^n}^{i_n} F = 0$ determines what is called a
{\it solution manifold}, which in general is an infinite-dimensional
submanifold of the jet space. In what follows we tacitly assume that
the total derivatives are restricted on the solution manifold;
these restrictions are tangent to the latter.

As usual, the {\it directional derivative} along an $N$-component vector
$\bi{U}=(U^1,\dots,U^N)^{T}$, where the superscript $T$ indicates the transposed matrix,
is the vector field
$$
\dder{\bi{U}} = \sum_{\alpha=1}^{N}\sum_{i_1,\dots,i_n=0}^{\infty}
(D_{x^1}^{i_1}\cdots D_{x^n}^{i_n} U^\alpha)
\frac{\p}{\p u^\alpha_{i_1\dots i_n}}.
$$
The total derivatives as well as
the directional derivative can be applied to (possibly vector or matrix) local functions
$P$.
%
Recall \cite{olv_eng2, vinbook} that $\bi{U}$ is a (characteristic of a)
{\it symmetry} for the system $F=0$ if
$\bi{U}$ satisfies $\dder{\bi U} F = 0$ on the solution manifold.

Assume now that the system $F=0$ can be written as a commutativity condition
$[X_1,X_2]=0$, where
$X_i = \sum_{j=1}^n \xi^j_i D_{x^j}$
are vector fields, $[\,\cdot\,,\cdot\,]$ is the usual Lie bracket thereof,
and the coefficients $\xi^j_i$ are local functions that may further depend
on a spectral parameter $\lambda$.

Further consider a vector field $Z$ of the same form, i.e., 
$Z=\sum_{j=1}^n \zeta^j D_{x^j}$,
except that we do not insist that $\zeta^j$ are local functions. However, we 
assume that the total derivatives can be extended to $\zeta^j$, see below.

The main idea of the present paper is that we
look for an $N \times n$ matrix $A = (a^\alpha_j)$
such that\looseness=-1
$$
\tilde {\bi U}^\alpha=\sum\limits_{j=1}^n
a^\alpha_j \zeta^j
$$
are components of a symmetry $\tilde{\bi U}$ whenever $Z$ satisfies
\be
\label{zcov}
[X_i,Z] = \dder{\bi{U}} X_i, \quad i=1,2.
\ee

We shall write
$\tilde{\bi U} = \mathfrak{R}_A(\bi{U})$ when such a matrix
$A = (a^\alpha_j)\not\equiv 0$ exists; this is precisely a recursion operator
of the type described in Introduction, i.e., a B\"acklund auto-transformation
for the linearized system $\dder{\bi U} F = 0$.
Here and below we assume for simplicity
that the entries $a^\alpha_j$ of $A$ are local functions; however, in principle,
nothing prevents them from being nonlocal.

Note that we do not insist that the vector fields $X_i$ necessarily
involve any spectral parameter,
but we do exclude the case when they involve derivatives with respect to the spectral
parameter.

The condition~(\ref{zcov}) is a system of first-order partial differential equations
in the unknowns $\zeta^j$.
To show that the system is compatible we check the Jacobi identity
$$
[X_1,[X_2,Z]] + [X_2,[Z,X_1]] + [Z,[X_1,X_2]]
 = [X_1,\dder{\bi U} X_2] - [X_2,\dder{\bi U} X_1]
 = \dder{\bi U} [X_1,X_2] = 0
$$
since $[X_1,X_2]=0$ is equivalent to $F = 0$ and $\bi U$ is a symmetry.

As a rule, the system~(\ref{zcov}) is not solvable in terms of local functions.
Therefore, strictly speaking, $\tilde {\bi U}$ are not necessarily symmetries of the
system $F = 0$.
Instead, they are {\em nonlocal} symmetries (or shadows in the sense of~\cite{K-V,vinbook}).
This naturally leads to introduction of pseudopotentials (for instance,
$\zeta^i$ and their derivatives) and subsequent extension of the total
derivatives to include the terms coming from pseudopotentials.
To simplify notation we shall, however, denote the extended
total derivatives by the same symbol $D_{x^i}$. Note that when
applied to local functions the original and extended total derivatives
coincide.\looseness=-1


Note that in a number of examples, where the recursion operators are already known,
e.g.\ the Pavlov equation \cite{P, M-S2}, our method produces the recursion
operators which are inverse to the known ones. Moreover, the inverses
of our recursion operators often have simpler structure
of nonlocal terms; in particular, this holds for all systems discussed
below. Thus, it is often appropriate to invert the operator $\mathfrak{R}_A$
resulting from the above construction in order to obtain a simpler recursion operator;
the inversion is an algorithmic process described in~\cite{Gut1}.

Let us also mention that, in sharp contrast with
the case of (1+1)-dimensional systems where one usually can make a clear
distinction among positive (local) and negative
(nonlocal) hierarchies (see, however, \cite{B}),
the multidimensional hierarchies we have been able to generate
contain, an eventual inversion of the recursion operator notwithstanding,
only a few local symmetries. The same phenomenon occurs for the multidimensional hierarchies
generated using bilocal recursion operators, see e.g.\ \cite{M-Y, Wang2}
and references therein. \looseness=-1

\looseness=-1

\section{The Manakov--Santini system}

Consider the Manakov--Santini system \cite{M-S} in $(N+4)$ independent variables
$x^1,\dots,x^N$, $y^1,y^2,z^1,z^2$ and $N$ dependent variables $u^i$,
\begin{equation}\label{mss}
u^i_{y^1z^2} - u^i_{y^2z^1} +
\sum_{j=1}^N (u^j_{z^1}u^i_{z^2 x^j} - u^j_{z^2}u^i_{z^1 x^j})
= 0,\quad i=1,\dots,N.
\end{equation}
As usual, the subscripts refer to partial derivatives.

System (\ref{mss}) can be written \cite{M-S} as a commutativity condition of the vector fields
\begin{equation}\label{laxmss}
X_i = D_{y^i} + \lambda D_{z^i} +
\sum_{k=1}^N u^k_{z^i} D_{x^k}, \quad i=1,2.
\end{equation}

Assume that $Z$ has the form
\begin{equation}
\label{zmss}
Z = \sum_{j=1}^N V^j D_{x^j},
\end{equation}
(no terms involving $D_{y^i}$ and $D_{z^k}$ are actually needed).

It is straightforward to verify that the following assertion holds for any natural $N$:
if $\boldsymbol{U} = (U^1,\dots,U^N)^T$ is a characteristic of symmetry
for (\ref{mss}) then so is $\boldsymbol{V} = (V^1,\dots,V^N)^T$, where
$V^j$ are determined from the equations (\ref{zcov}) with $X_i$ and $Z$ given by
(\ref{laxmss}) and (\ref{zmss}), that is,
\begin{equation}
\label{zcovmss}
V^i_{y^s} + \lambda V^i_{z^s}
 + \sum_{j=1}^N u^j_{z^s} V^i_{x^j}
= \sum_{j=1}^N u^i_{z^s x^j} V^j
 - U^i_{z^s},
\quad i=1,\dots,N, \quad s=1,2.
\end{equation}

To emphasize the dependence on $\lambda$, the recursion operator given by
formula~(\ref{zcovmss}) will be denoted $\mathfrak R_\lambda$.
Applying $\mathfrak R_\lambda$ to local symmetries yields
a highly nonlocal `negative' hierarchy of the Manakov--Santini system.
In order to obtain the `positive' hierarchy with simpler
nonlocalities, we look for the
inverse~$\mathfrak R_\lambda^{-1}$.

Inverting the recursion operator $\mathfrak R_\lambda$ amounts to solving (\ref{zcovmss})
for~$U^i_{z^j}$.
The inverse operator $\mathfrak R_\lambda^{-1}$ sends
$\boldsymbol{V}=(V^1,\dots,V^N)^T$ to $\boldsymbol{U} = (U^1,\dots,U^N)^T$, where
$U^i$ are determined from the relations
$$
U^i_{z^s} = -V^i_{y^s}
 - \lambda V^i_{z^s}
 - \sum_{j=1}^N u^j_{z^s} V^i_{x^j}
 + \sum_{j=1}^N u^i_{z^s x^j} V^j,
\quad i=1,\dots,N, \quad s=1,2.
$$
Upon multiplying by $-1$ and removing the trivial contribution $\lambda \boldsymbol{V}$
from $\boldsymbol{U}$,
we end up with the recursion operator
$\mathfrak R = -\mathfrak R_\lambda^{-1} - \lambda\,\mathrm{Id}$,
which no longer depends on $\lambda$.
The components of
$\bi U = \mathfrak R(\bi V)$ are defined by the relations
\begin{equation}\label{mss RO}
U^i_{z^s} = V^i_{y^s}
 + \sum_{j=1}^N u^j_{z^s} V^i_{x^j}
 - \sum_{j=1}^N u^i_{z^s x^j} V^j,
\quad i=1,\dots,N, \quad s=1,2.
\end{equation}
The symmetries generated using this recursion operator are complicated nonlocal expressions;
the explicit form for symmetries obtained by applying (\ref{mss RO}) to the Lie point symmetries
of (\ref{mss}) is given in the appendix.

\section{Dispersionless Hirota equation}

Consider the equation \cite{zakharevich,bft}
\begin{equation}\label{dhe}
a u_x u_{yt} + b u_y u_{xt} + c u_t u_{xy} = 0,\quad a+b+c=0.
\end{equation}
It has a Lax pair \cite{zakharevich} of the form
\[
\psi_y=\frac{\lambda u_y\psi_x}{u_x},\quad \psi_t=\frac{\mu u_t\psi_x}{u_x},
\]
where $\mu=(a+b)\lambda/(a\lambda+b)$.

The vector field $Z$ now can be chosen in the form
$Z = \zeta D_x$.
An easy computation shows that the corresponding recursion operator is given by the formula
$$
\mathfrak{R}_\lambda(U)=u_x \zeta.
$$
Here $U$ is a symmetry for (\ref{dhe}) and $\zeta$ is defined by the following equations:
$$
\begin{array}{l}
\displaystyle
\frac{\zeta_t}{\mu} =
 \frac{u_t}{u_x} \zeta_x
 - \left(\frac{u_t}{u_x}\right)_x \zeta
 + \frac{u_t U_x - u_x U_t}{u_x^2},\quad 
\displaystyle
\frac{\zeta_y}{\lambda} =
 \frac{u_y}{u_x} \zeta_x
 - \left(\frac{u_y}{u_x}\right)_x \zeta
 + \frac{u_y U_x - u_x U_y}{u_x^2}.
 \end{array}
$$

The inverse recursion operator $W=\mathfrak{R}_\lambda^{-1}(U)$ is given by the formulas
\[
\ba{l}\displaystyle
W_y = \frac{u_y W_x}{u_x}
 + \frac{u_y U_x}{u_x}
 - \frac{U_y}{\lambda}
 - \frac{\lambda - 1}\lambda \frac{u_{xy}}{u_x} U,\quad
\displaystyle
W_t = \frac{u_t W_x}{u_x}
 + \frac{u_t U_x}{u_x}
 - \frac{U_t}{\mu}
 - \frac{\mu - 1}{\mu} \frac{u_{xt}}{u_x} U.
\ea
\]
If we replace $W$ by $W - U$, we obtain a somewhat simpler recursion operator $W = \tilde{\mathfrak{R}}(U)$, where
$\tilde{\mathfrak{R}} = \mathfrak{R}_\lambda^{-1} + \mathrm{Id}$,
so $W$ can be determined from the compatible equations
$$
\begin{array}{l}
\displaystyle
W_y = \frac{u_y}{u_x} W_x
 + \frac{\lambda - 1}\lambda \left(U_y - \frac{u_{xy}}{u_x} U\right),\quad 
\displaystyle
W_t = \frac{u_t}{u_x} W_x
 + \frac{\mu - 1}\mu \left(U_t - \frac{u_{xt}}{u_x} U\right).
\end{array}
$$

Let us apply $\tilde{\mathfrak{R}}$ to the Lie point symmetries, which are $X(x) u_x$, $Y(y) u_y$, $T(t) u_t$,
and $F(u)$.
To start with, we find $\tilde{\mathfrak{R}}(0) = F(u)$, where $F$ is an arbitrary smooth function.
Upon having agreed to remove this trivial contribution from the results, we readily find
$$
\tilde{\mathfrak{R}}(X u_x) = 0, \quad
\tilde{\mathfrak{R}}(Y u_y) = \frac{\lambda - 1}\lambda Y u_y, \quad
\tilde{\mathfrak{R}}(T u_t) = \frac{\mu - 1}\mu T u_t.
$$
Thus, $X u_x, Y u_y, T u_t$ are eigenvectors of $\tilde{\mathfrak{R}}$. To the best
of our knowldege, this is a first known example of {\em local\/} eigenvectors for
a nontrivial recursion operator.

Finally,
$$
\tilde{\mathfrak{R}}(F(u)) =
F p + \frac{\partial F}{\partial u} q,
$$
where $p,q$ are nonlocal variables defined by the following equations which are
compatible by virtue of (\ref{dhe}):\looseness=-1
\[
\begin{array}{l}
\displaystyle p_y = \frac{u_y}{u_x} p_x - \frac{\lambda-1}{\lambda} \frac{u_{xy}}{u_x},
\quad
p_t = \frac{u_t}{u_x} p_x - \frac{\mu-1}{\mu} \frac{u_{xt}}{u_x},\quad 
\displaystyle q_y = \frac{u_y}{u_x} q_x + \frac{\lambda-1}{\lambda} u_y,
\quad
q_t = \frac{u_t}{u_x} q_x - \frac{\mu-1}{\mu} u_t.
\end{array}
\]

\section{The case of Hamiltonian vector fields}

If the Lax pair for the system under study consists
of Hamiltonian vector fields,
it is natural to apply the ideas from Section~\ref{ga}
to the algebra of functions endowed with the Poisson bracket
rather than to the algebra of vector fields.

Namely, suppose that the Lax pair for the system under study
can (up to the obvious renumbering of independent variables) be written as
\begin{equation}\label{laxpb}
\psi_{x^{n-1}}=a\psi_{x^{n-2}}+\{H_1,\psi\},\quad
\psi_{x^{n}}=b\psi_{x^{n-3}}+\{H_2,\psi\}.
\end{equation}
Here $\{\cdot,\cdot\}$ denotes the Poisson bracket in question (usually w.r.t.\ the independent
variables $x^1,\dots,\allowbreak x^{n-4}$ only), and $a,b$ are some constants,
which are typically proportional to the spectral parameter $\lambda$.\looseness=-1

Instead of $\zeta^j$ introduce a single
nonlocal variable $\Omega$ defined by the formulas
\begin{equation}\label{omega}
\Omega_{x^{n-1}}=a\Omega_{x^{n-2}}+\{H_1,\Omega\}+\dder{\bi{U}} H_1,\quad
\Omega_{x^{n}}=b\Omega_{x^{n-3}}+\{H_2,\Omega\}+\dder{\bi{U}} H_2,
\end{equation}
where $a,b$ are some constants which are often proportional to the spectral parameter $\lambda$.

Then we shall seek for a recursion operator in the form
\begin{equation}\label{roham}
\tilde{\bi U} = \mathfrak{R}(\bi{U}) = \boldsymbol{A}_0 \Omega +\sum\limits_{i=1}^{n-2} \boldsymbol{A}_i \Omega_{x^i}.
\end{equation}
where now $\boldsymbol{A}_j= (a_j^1,\dots,a_j^N)$, $j=0,\dots,n-2$, are $N$-component vectors whose entries are local functions. If necessary, the terms containing higher-order derivatives can be also included.

As an example, consider the following
six-dimensional generalization of the first heavenly equation, see e.g.\ \cite{df,pp}:
\begin{equation}\label{6d_fhe}
u_{xp}=-u_{yq}+u_{xt} u_{yz}-u_{xz}u_{yt}.
\end{equation}
It admits \cite{df} a Lax representation of the form (\ref{laxpb}), namely,
\[
\psi_{p}=\lambda\psi_y+\{u_y,\psi\},\quad
\psi_{q}=-\lambda\psi_x+\{-u_x,\psi\},
\]
with the Poisson bracket given by
\[
\{f,g\}=D_z (f) D_t(g)-D_t (f) D_z(g).
\]

It is readily verified that (\ref{6d_fhe}) possesses a recursion operator of the form
$\tilde U\equiv\mathfrak{R}(U)=\Omega$, where the nonlocal variable $\Omega$ is defined via (\ref{omega}), that is,
\[
\ba{l}
\Omega_{p}=\lambda\Omega_{y}+\{u_y,\Omega\}+U_y\equiv \lambda\Omega_{y}+u_{yz}\Omega_t-u_{yt}\Omega_z+U_y,\\[3mm]
\Omega_{q}=-\lambda\Omega_{x}+\{-u_x,\Omega\}-U_x\equiv -\lambda\Omega_{x}-u_{xz}\Omega_t+u_{xt}\Omega_z-U_x.
\ea
\]
Upon inversion we obtain a simpler recursion operator $\widetilde{U}=(\mathfrak{R}^{-1}+\lambda\ \mathrm{Id})(U)$, where $\widetilde{U}$ is defined by the formulas
\begin{equation}\label{6d_fhe_iro}
\ba{l}
\widetilde{U}_x=-u_{xz} U_t+u_{xt}U_z-U_q,\quad
\widetilde{U}_y=-u_{yz}U_t+u_{yt}U_z+U_p.
\ea
\end{equation}

For another example, consider the modified heavenly equation \cite{df},
\begin{equation}\label{mhe}
u_{xz}=u_{xy} u_{tt}-u_{xt}u_{yt},
\end{equation}
which has \cite{df} a Lax representation of the form (\ref{laxpb}),
\[
\psi_{x}=\{u_x/\lambda,\psi\},\quad
\psi_{z}=\lambda\psi_t+\{-u_t,\psi\},
\]
with the Poisson bracket given by
\[
\{f,g\}=D_y (f) D_t(g)-D_t (f) D_y(g).
\]
It is readily seen that (\ref{mhe}) admits a recursion operator
$\tilde U\equiv\mathfrak{R}(U)=\Omega$, where the nonlocal variable $\Omega$ is now
defined by the formulas
\[
\ba{l}
\Omega_{x}=\{u_x/\lambda,\Omega\}+U_x/\lambda\equiv u_{xy}\Omega_t/\lambda-u_{xt}\Omega_y/\lambda+U_x/\lambda,\\[3mm]
\Omega_{z}=\lambda\Omega_{t}+\{-u_t,\Omega\}-U_t\equiv \lambda\Omega_{t}+u_{tt}\Omega_y-u_{yt}\Omega_t-U_t.
\ea
\]
Inversion again leads to a simpler recursion operator $\widetilde{U}=(\mathfrak{R}^{-1}-\lambda\ \mathrm{Id})(U)$,
with $\widetilde{U}$ defined by the formulas
\begin{equation}\label{mhe_iro}
\ba{l}
\widetilde{U}_{x}=u_{xt}U_y-u_{xy}U_t,\quad 
\widetilde{U}_{t}=u_{tt}U_y-u_{yt}U_t-U_z.
\ea
\end{equation}

To the best of our knowledge, the recursion operator (\ref{6d_fhe_iro}) has not yet appeared in the literature, while (\ref{mhe_iro}) is a special case of the recursion operator for the so-called asymmetric heavenly equation found in \cite{She} using the partner symmetry approach.
Note that (\ref{6d_fhe_iro}) also could have been obtained within the partner symmetry approach \cite{She}. On the other hand, the recursion operators for the second heavenly and Husain equations, which were
found in \cite{ney,shef1}, can be easily recovered using the approach of the present section.
\looseness=-1

\subsection*{Acknowledgements}
A.S. gratefully acknowledges the discussion of the results of the present paper with J.D.E. Grant and with B.G. Konopelchenko. A.S. also thanks J.D.E. Grant for bringing the reference \cite{GS} to his attention and B.G. Konopelchenko for pointing out the reference \cite{B-K} and the fact that auxiliary linear problems involving commuting vector fields were considered for the first time by Zakharov and Shabat in \cite{Z-S}.

This research was supported in part by the
Ministry of Education, Youth and Sports of Czech Republic (M\v SMT \v CR) under grant
MSM4781305904, and by the Czech Grant Agency (GA \v CR) under
grant P201/11/0356.

\appendix

\section*{Appendix: Symmetries of the Manakov--Santini system}

In this section we use the standard convention on summation over repeated indices. The
indices $r,s$ run from $1$ to $2$, the others run from $1$ to $n$.

The symmetries of the Manakov--Santini system can be routinely computed as
solutions $\bi U$ of the linearized equation $\dder{\bi U} F = 0$ which must
hold only on the solution manifold of  (\ref{mss}); here
$F^i$ stands for the left-hand side of $i$th equation of (\ref{mss}).

The simplest of symmetries, the Lie point ones, are characterized by the property that $U^i$
are linear in the first derivatives.
A computer-aided computation reveals fourteen Lie point symmetries for (\ref{mss}), namely,
$$
\eta_1,\eta_2,\zeta_1,\zeta_2,\alpha,
\beta_{11},\beta_{12},\beta_{21},\beta_{22},
\gamma,\chi_1,\chi_2,{} _f\xi,{} _g\phi,
$$
where ${} _f\xi,\ _g\phi$ each depend on $N$
arbitrary functions 
of the coordinates $x^i,y^j$, $i,j=1,\dots,N$; the left-hand-side subscripts indicate the arbitrary
functions these symmetries depend on.

The generators of these Lie point symmetries read
$$
\begin{array}{l}
\eta^k_r = u^k_{y^r},\quad 
\zeta^k_r = u^k_{z^r},\quad 
\alpha^k = y^s u^k_{z^s},\quad 
(\beta^r_s)^k = y^r u^k_{y^s} + z^r u^k_{z^s},\quad 
\gamma^k = u^k + y^s u^k_{y^s}, \\[3mm]
(\chi^r)^k = y^r u^k
      + y^r y^s u^k_{y^s} + z^r y^s u^k_{z^s},\quad 
_f\xi^k = f^k(x,y),\quad 
_g\phi^k =\displaystyle -g^j(x,y) u^k_{x^j}
 + \frac{\partial g^k(x,y)}{\partial x^j} u^j
 + \frac{\partial g^k(x,y)}{\partial y^s} z^s.
\end{array}
$$
This notation is to be read as follows: for instance,
the symmetry $_g\phi$ has a characteristic $(_g\phi^1,\dots,\ _g\phi^N)$
and the associated evolutionary vector
field is $\sum_{\nu=1}^N {} _g\phi^\nu\partial/\partial u^\nu$.

Let us investigate the action of the recursion operator (\ref{mss RO})
on the above symmetries.
Obviously, equations~\eqref{mss RO} determine $U^k$
uniquely up to adding the Lie symmetry we denoted $_f\xi$
(we could also write $_f\xi = \mathfrak R(0)$).
Like the integration constants, this term will
be omitted in what follows.


Four classical symmetries, namely
$\xi, \zeta_1, \zeta_2, \alpha$, are mapped to local symmetries again:
$$
\mathfrak R(_f\xi) ={} _f\phi, \quad
\mathfrak R(\zeta_r) = \eta_r, \quad
\mathfrak R(\alpha) = \gamma.
$$
The others are mapped to nonlocal symmetries, sharing the same set of nonlocal
variables $u^{(2)k}$, $k = 1,\dots,N$, subject to the equations
\be
\label{MS u(2)}
u^{(2)k}_{z^r} = u^k_{y^r} + \sum_{i = 1}^N u^i_{z^r} u^k_{x^i}.
\ee

The system~\eqref{MS u(2)} is compatible by virtue of the
Manakov--Santini system (\ref{mss}). Eq.(\ref{MS u(2)}) determines a
covering in the sense of~\cite{vinbook},
but this covering is infinite-dimensional.
Each successive application of $\mathfrak R$ requires one more level of nonlocal
variables $u^{(q)k}$, subject to the compatible equations
\be
\label{MS u(q)}
u^{(q+1) k}_{z^r} = u^{(q) k}_{y^r} + \sum_{i = 1}^N u^i_{z^r} u^{(q) k}_{x^i}.
\ee
Thus we have obtained an infinite hierarchy of successive coverings.

Upon denoting $\eta^{(q)}_r\equiv \mathfrak R^q(\eta_r)
 = \mathfrak R^{(q+1)}(\zeta_r)\equiv \zeta^{(q+1)}_r$,
we routinely generate
$$
\begin{array}{l}
\eta_r^{(1) k} = \zeta_r^{(2) k}
        = u^{(2)k}_{y^r}
        - u^k_{x^i} u^i_{y^r} , \\[3mm]
\eta_r^{(2) k} = \zeta_r^{(3) k}
        = u^{(3)k}_{y^r}
        - u^{(2)k}_{x^i} u^i_{y^r}
        - u^k_{x^i} u^{(2)i}_{y^r}
        + u^k_{x^i} u^i_{x^j} u^j_{y^r}
         , \\[3mm]
\eta_r^{(3) k} = \zeta_r^{(4) k}
        = u^{(4)k}_{y^r}
        - u^{(3)k}_{x^i} u^i_{y^r}
        - u^{(2)k}_{x^i} u^{(2)i}_{y^r}
        - u^k_{x^i} u^{(3)i}_{y^r}
\\[3mm]\qquad
        + u^{(2)k}_{x^i} u^i_{x^j} u^j_{y^r}
        + u^k_{x^i} u^{(2)i}_{x^j} u^j_{y^r}
        + u^k_{x^i} u^i_{x^j} u^{(2)j}_{y^r}
        - u^k_{x^i} u^i_{x^j} u^j_{x^h} u^h_{y^r},
        \end{array}
$$
etc. If we assign level one to the local variables, i.e., $u^k = u^{(1)k}$,
we observe that
the sum in the above formulas runs over all homogeneous monomials of the same level.
We conjecture that this pattern holds for $\eta^{(q)}_r$ with any natural $q$.

In terms of these, we observe the following general formula for
$\beta^{(q)}\equiv \mathfrak R^q(\beta)$:
$$
(\beta^{(q) r}_s)^k
  = y^r \eta_s^{(q) k} + z^r \zeta^{(q) k}_s.
$$
Likewise, for $\gamma^{(q)}\equiv\mathfrak R^q(\gamma)
 = \mathfrak R^{q+1}(\alpha) \equiv\alpha^{(q+1)}$ we obtain the expressions
$$
\begin{array}{l}
\gamma^{(1) k} = \alpha^{(2) k} = 2 u^{(2) k}
       - u^i u^k_{x^i}
       + y^s \eta^{(1) k}_s,\allowbreak \\[3mm]
\gamma^{(2) k} = \alpha^{(3) k} = 3 u^{(3) k}
        - u^{(2)k}_{x^i} u^i
        - 2 u^k_{x^i} u^{(2)i}
        + u^k_{x^i} u^i_{x^j} u^j
        + y^s \eta^{(2) k}_s,\allowbreak \\[3mm]
\gamma^{(3) k} = \alpha^{(4) k} = 4 u^{(4) k}
        - u^{(3)k}_{x^i} u^i
        - 2 u^{(2)k}_{x^i} u^{(2)i}
        - 3 u^k_{x^i} u^{(3)i}
\\[3mm]\qquad
        + u^{(2)k}_{x^i} u^i_{x^j} u^j
        + u^k_{x^i} u^{(2)i}_{x^j} u^j
        + 2 u^k_{x^i} u^i_{x^j} u^{(2)j}
        - u^k_{x^i} u^i_{x^j} u^j_{x^h} u^h
        + y^s \eta^{(3) k}_s,\allowbreak \\[3mm]
\gamma^{(4) k} = \alpha^{(5) k} = 5 u^{(5) k}
        - u^{(4)k}_{x^i} u^i
        - 2 u^{(3)k}_{x^i} u^{(2)i}
        - 3 u^{(2)k}_{x^i} u^{(3)i}
        - 4 u^k_{x^i} u^{(4)i}
        + u^{(3)k}_{x^i} u^{i}_{x^j} u^{j}
        + u^{(2)k}_{x^i} u^{(2)i}_{x^j} u^{j}
\\[3mm]\qquad         
        + 2 u^{(2)k}_{x^i} u^{i}_{x^j} u^{(2)j}       
        + u^{k}_{x^i} u^{(3)i}_{x^j} u^{j}
        + 2 u^{k}_{x^i} u^{(2)i}_{x^j} u^{(2)j}
        + 3 u^{k}_{x^i} u^{i}_{x^j} u^{(3)j}
        - u^{(2)k}_{x^i} u^{i}_{x^j} u^{j}_{x^g} u^{g}
        - u^{k}_{x^i} u^{(2)i}_{x^j} u^{j}_{x^g} u^{g}
\\[3mm]\qquad        
        - u^{k}_{x^i} u^{i}_{x^j} u^{(2)j}_{x^g} u^{g}
        - 2 u^{k}_{x^i} u^{i}_{x^j} u^{j}_{x^g} u^{(2)g}
        + u^{k}_{x^i} u^{i}_{x^j} u^{j}_{x^g} u^{g}_{x^h} u^{h}
        + y^s \eta^{(4) k}_s.
\end{array}
$$
Here another pattern can be observed. Namely,
the sum again runs over all homogeneous monomials with the coefficient at each monomial
equal, up to the sign, to the level of the nonlocal variable which is not differentiated. It would be interesting to find out whether
this pattern holds for all natural $q$.

In terms of $\gamma^{(q)}$ we have the following general formula for
$\chi^{(q)}\equiv \mathfrak R^q(\chi)$:
$$
(\chi^r)^{(q) k}=y^r \gamma^{(q) k} + z^r \alpha^{(q) k}.
$$
Finally, we have $_g\phi^{(q)}\equiv \mathfrak R^q(_g\phi)
 = \mathfrak R^{(q+1)}(_g\xi)\equiv{} _g\xi^{(q+1)}$,
$$
\begin{array}{l}\displaystyle
_g\phi^k =\displaystyle _g\xi^{(1) k}= -u^k_{x^i} g^i
 + \frac{\partial g^k}{\partial x^i} u^i
 + \frac{\partial g^k}{\partial y^s} z^s,
\\[3mm]
_g\phi^{(1) k} =\displaystyle _g\xi^{(2) k} = -u^{(2)k}_{x^i} g^i
       + \frac{\partial g^k}{\partial x^i}u^{(2)i}
       + u^k_{x^j} \left(-u^j_{x^i} g^i
                  + \frac{\partial g^j}{\partial x^i} u^i
                  + \frac{\partial g^j}{\partial y^s} z^s\right)  \\[5mm]
\qquad \displaystyle+ \frac12 \left(\frac{\partial^2 g^k}
{\partial x^i\,\partial x^j} u^i u^j
       + 2 \frac{\partial g^k}{\partial x^i\,\partial y^s} u^i z^s
       + \frac{\partial^2 g^k}{\partial y^r\,\partial y^s} z^r z^s\right).
\end{array}
$$

\end{document}